\documentclass[sn-mathphys,Numbered]{sn-jnl}

\usepackage{tabularray}
\usepackage{pdflscape} 
\usepackage{xr}
\usepackage{graphicx}%
\usepackage{multirow}%
\usepackage{amsmath,amssymb,amsfonts}%
\usepackage{amsthm}%
\usepackage{mathrsfs}%
\usepackage[title]{appendix}%
\usepackage{xcolor}%
\usepackage{textcomp}%
\usepackage{manyfoot}%
\usepackage{booktabs}%
\usepackage{algorithm}%
\usepackage{algorithmicx}%
\usepackage{algpseudocode}%
\usepackage{listings}%
\usepackage{url}
\usepackage{soul}
\usepackage{enumitem}
\usepackage{tikz}
\usepackage{circledtext}
\usepackage{enumitem}
\usepackage{circledsteps}
\usepackage{tabularray}
\usepackage{pdflscape} 
\usepackage{amsmath}

\usepackage[printonlyused]{acronym}
\usepackage{makecell}
\usepackage[normalem]{ulem}

\acrodef{FoV}{field-of-view}
\acrodef{UAV}{Unmanned Aerial Vehicles} 
\acrodef{AoA}{Angle-of-Arrival}
\acrodef{IoT}{Internet-of-Things}
\acrodef{LEO}{Low-Earth Orbit}
\acrodef{SNR}{signal-to-noise ratio}
\acrodef{RX}{receiver}
\acrodef{FPGA}{field-programmable gate array}
\acrodef{TIA}{trans-impedance amplifier}
\acrodef{LNTA}{low-noise transconductance amplifier}
\acrodef{BFIC}{Beamforming Integrated Circuit}
\acrodef{REF}{Reference}
\acrodef{LO}{Local Oscillator}
\acrodef{HPF}{high-pass filter}
\acrodef{PI}{phase-interpolator}
\acrodef{PS}{phase-shifter}
\acrodef{BUF}{buffer}
\acrodef{OTA}{Operational trans-conductance amplifier}
\acrodef{LUT}{look-up table}
\acrodef{3D}{three-dimensional}
\acrodef{DBS}{Dynamic Beam-Stabilized}
\acrodef{SWaP-C}{size, weight, area, power, and cost}
\acrodef{PLL}{Phase Locked Loop}
\acrodef{DAC}{Digital to Analog Converter}
 \acrodef{CuMOD}{Copper molecular decomposition }
 \acrodef{TPU}{Thermoplastic polyurethane}
\acrodef{PET}{Polyethylene terephthalate}
\acrodef{TGA}{Thermogravimetric analysis}
\acrodef{SWaP-C}{size, weight, area, power, and cost}
\acrodef{PLL}{Phase Locked Loop}
\acrodef{DAC}{Digital-to-Analog Converter}
\acrodef{ADC}{Analog-to-Digital Converter}
\acrodef{TPU}{Thermoplastic polyurethane}
\acrodef{PET}{Polyethylene terephthalate}
\acrodef{TGA}{Thermogravimetric analysis}
 \acrodef{CuMOD}{Copper Molecular Decomposition }
 \acrodef{EMI}{electromagnetic interference}
 \acrodef{RF}{Radio Frequency}

\definecolor{deepblack}{rgb}{0,0,0.8}
\definecolor{deepred}{rgb}{0.6,0,0}
\definecolor{deepgreen}{rgb}{0,0.5,0}
\definecolor{backcolor}{rgb}{0.95,0.95,0.92}
\definecolor{light-gray}{HTML}{E5E4E2}
\definecolor{light-cyan}{HTML}{E0FFFF}
\definecolor{sage}{HTML}{EEF1DA}
\definecolor{light-teal}{HTML}{C7D9DD}

\theoremstyle{thmstyleone}%
\theoremstyle{thmstyletwo}%

\theoremstyle{thmstylethree}%
\begin{document}

\title[Article Title]{Dynamic Beam-Stabilized,  Additive-Printed Flexible Antenna Arrays with On-Chip \textcolor{black}{Rapid} Insight Generation}


\author*[1]{\fnm{Sreeni} \sur{Poolakkal}}\email{sreeni.poolakkal@wsu.edu}
\author[2]{\fnm{Abdullah} \sur{Islam}}
\author[1]{\fnm{Arpit} \sur{Rao}}
\author[1]{\fnm{Shrestha} \sur{Bansal}}
\author[3]{\fnm{Ted} \sur{Dabrowski}}
\author[3]{\fnm{Kalsi} \sur{Kwan}}
\author[2]{\fnm{Zhongxuan} \sur{Wang}}
\author[4]{\fnm{Amit Kumar} \sur{Mishra}}
\author[3]{\fnm{Julio} \sur{Navarro}}
\author[2]{\fnm{Shenqiang} \sur{Ren}}
\author[3]{\fnm{John} \sur{Williams}}
\author[4]{\fnm{Sudip} \sur{Shekhar}}
\author[1]{\fnm{Subhanshu} \sur{Gupta}}

\affil[1]{\orgdiv{School of Electrical Engineering and Computer Sciences}, \orgname{Washington State University}, \orgaddress{\street{355 NE Spokane St}, \city{Pullman}, \postcode{99163}, \state{WA}, \country{USA}}}

\affil[2]{\orgdiv{Department of Materials Science and Engineering}, \orgname{University of Maryland}, \orgaddress{ \state{MD}, \country{USA}}}

\affil[3]{\orgdiv{Additive Printing}, \orgname{Boeing}, \orgaddress{ \state{AL}, \country{USA}}}

\affil[4]{\orgdiv{Department of Electrical and Computer Engineering}, \orgname{University of British Columbia}, \orgaddress{ \state{BC}, \country{Canada}}}


\abstract{

Conformal phased arrays promise shape-changing properties, multiple degrees of freedom to the scan angle, and novel applications in wearables, aerospace, defense, vehicles, and ships. However, they have suffered from two critical limitations. (1) Although most applications require on-the-move communication and sensing, prior conformal arrays have suffered from dynamic deformation-induced beam pointing errors. We introduce a \ac{DBS} processor capable of beam adaptation through on-chip real-time control of fundamental gain, phase, and delay for each element. (2) Prior conformal arrays have leveraged additive printing to enhance flexibility, but conventional printable inks based on silver are expensive, and those based on copper suffer from spontaneous metal oxidation that alters trace impedance and degrades beamforming performance. We instead leverage 
a low-cost \ac{CuMOD} ink with $<0.1\%$ variation per $^\circ$C with temperature and strain and correct any residual deformity in real-time using the \ac{DBS} processor. Demonstrating unified material and physical deformation correction, our CMOS DBS processor is low-power, low-area, and easily scalable due to a tile architecture, thereby ideal for on-device implementations.  

}  




\maketitle
\section{Introduction}
\label{sec:intro}

\begin{figure}[b!]
\centering
\vspace{-3mm}
\includegraphics[width=125mm]{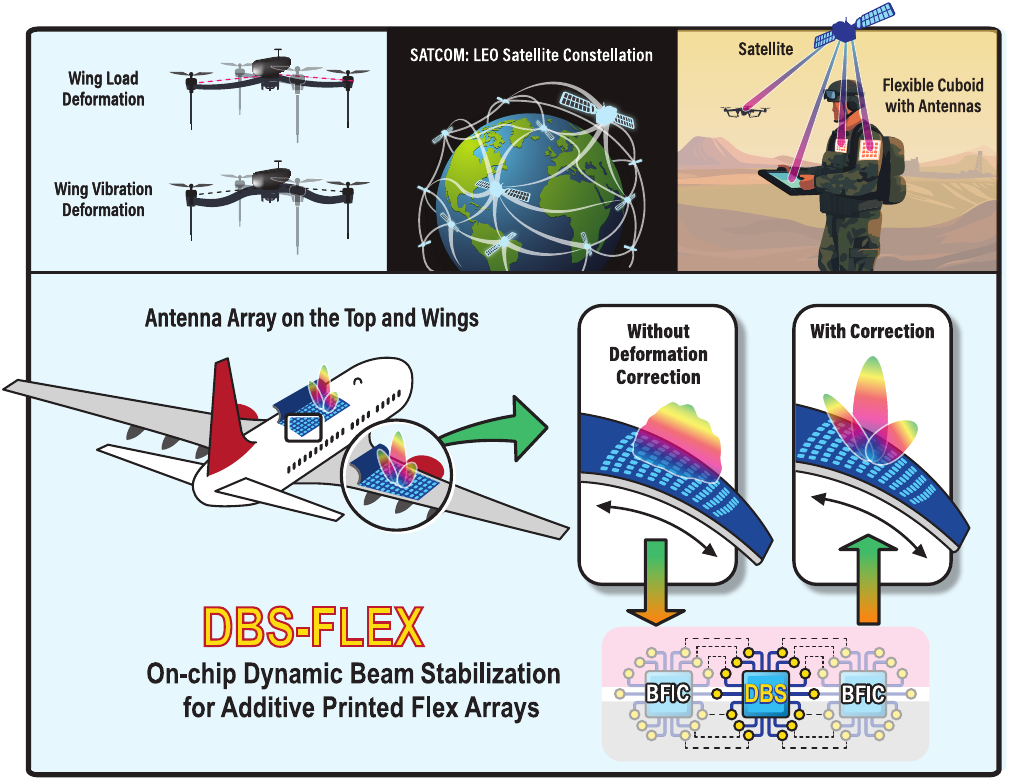}
\caption{\small (Top left) Dynamic wing load variations and wing vibrations lead to deformed antenna arrays impacting in-flight wireless and navigation in \ac{UAV}. Dynamic deformation in (top center) satellites and (top right) lightweight textile arrays. (Bottom) Our silicon-based \ac{DBS} processor addresses gain loss and beam pointing error in the Beamforming Integrated Circuit (BFIC) due to dynamic deformation in real-time.}
\vspace{-2mm}
\label{fig:Mot}
\end{figure}

Conformal phased antenna arrays
enabling added degrees of freedom in scan angle, thus the wider field of view is critical for several imaging radar and communication applications in the automotive, aviation, and space industries. The lightweight, small volume, and shape-changing properties~\cite{You_Access23,Fikes_Frontiers,matan_njp22} are attractive not only for structural applications and disaster relief management but also for wearable and textile arrays ~\cite{textile_1,textile_2,textile_3}. 

The promise of flexible arrays, however, is tempered by the use of conventional subtractive fabrication techniques that cause a decline in electrical and \ac{RF} performance under deformation due to the inflexibility of bulk copper traces~\cite{on_the_bending}.  At scale and volume, this incurs significant material wastage that is not environmentally or commercially sustainable. 
In contrast, additive printing has emerged as a promising alternative, enabling fabrication in a three-dimensional plane \cite{3d_permitivity}. Printable conductive inks used in the additive printing process not only provide more flexibility than the bulk copper~\cite{on_the_bending}, but also enable precise and rapid control of relative permittivity in all three axes, maintaining structural integrity and allowing fabrication of electromagnetic (EM) structures with complex geometry~\cite{3d_permitivity}. 
\textcolor{black}{Despite} its promises, additively printed active antenna arrays have been limited by material and physical deformities. State-of-the-art silver-based additive inks have been shown to exhibit 14~$\times$ higher resistivity than bulk silver or copper ($=220 n\Omega.cm$)~\cite{on_the_bending}. In ~\cite{inK_s11_change}, joule heating effects on trace resistivity and printing imperfections have been shown to affect electrical properties such as characteristic impedance and signal reflection coefficients. Further, the change in the phase constant necessitates additional phase compensation\footnote{Ink stability, trace impedance, and phase constant vary under strain and temperature and are impacted by ink composition, printing materials, printing technique, and strain together with the temperature variations. Silver-based ink varies between 0.1\% to 0.5\%  per degree Celsius with a gauge factor$ ~\approx2$~\cite{silver_ink_variation}, and Copper-based inks show a variation of $0.4\%$ to $0.6\%$ per degree Celsius and a similar gauge factor~\cite{copper_ink_variation}. Gauge factor is the percentage of resistance change with respect to strain. More details are in  Supplementary~1.6.}\textsuperscript{,}\footnote{Printing imperfections such as inconsistent deposition density (or dielectric constant) of the traces on the flexible substrates alters signal propagation speed as well as relative shifts in the positions of the antenna elements leading to change in the phase compensation at each element of the array.}. 
\textcolor{black}{ \textcolor{black}{ Furthermore, the surface deformation of a conformal phased array with curvature enhances its \ac{FoV} but increases inter-element coupling \cite{book_ch10} and introduces path length variations, leading to beam pointing errors \cite{book2}. The beam stabilization under static deformation, where inter-element coupling and path length variations are known a priori, is relatively straightforward to create a new beam steering codebook that accounts for the deformation effects~\cite{TAP23_DL,self_calib_TAB13}. While the initial (static) deformation due to the aerodynamic shape of aircraft and drones can be estimated, the dynamic deformations arising from wing loading and vibrations~\cite{TAP23_DL} during transit cannot be predetermined. These dynamic deformations depend on factors such as array weight, size, drone/aircraft speed, wind conditions, initial deformation, etc. \newline 
\indent State-of-the-art beam stabilization techniques are inadequate with unknown dynamic deformation correction, especially on portable applications such as \ac{UAV} and  aircraft~\cite{TAP23_DL,self_calib_TAB13,Hajimiri_JSSC21,self_calib_TMTT21,self_calib_IMS19}. }} \textcolor{black}{In~\cite{Hajimiri_JSSC21}, a multi-dimensional variable dynamic range search algorithm adjusts the array weights to stabilize the beam, but its higher computational complexity and parallelism demands costly \ac{FPGA} architectures, besides the additional requirement for separate generation and recovery units connected through a separate wireless back propagation link.  Further, phase correction using iterative methods such as genetic algorithm~\cite{Genetic_algorithm} and iterative phase synthesis ~\cite{array_synthesis1,array_synthesis2,iterative1} have leveraged complex \ac{FPGA} requiring significantly higher computational workloads, thereby incurring higher latencies not suited for low \ac{SWaP-C} applications. Adapting these techniques in airborne platoons and portable units is thus challenging. 
The pre-trained deep-learning-based beam stabilization technique presented in~\cite{TAP23_DL} relies heavily on training data with prior knowledge of the deformation, deformation surface, and operational environment. Under uncertain dynamic operational conditions, this technique fails to stabilize the beam. 
Furthermore, these models require periodic update and maintenance of multiple dictionary sets, demanding an exponential increase in memory storage as the array size is scaled.  Element-level mutual coupling-based deformation measurement and compensation using an off-chip algorithm was demonstrated in~\cite{Mizrahi_IMS21,self_calib_TMTT21}. The additional interfacing challenges for capturing element-level deformation and the additional computational requirements of the algorithm limit its use in dynamic deformation beam-stabilization scenarios. The element-level deformation sensing using mechanical strain sensors, as presented in~\cite{self_calib_TAB13}, faces similar interfacing challenges when used for dynamic beam-stabilization. This technique lacks the capability for automatic correction and can only be considered a calibration technique (see Supplementary~1.4 for comparative analysis with state-of-the-art). }
Recognizing such limitations, our work integrates advanced beamformer ICs with lightweight and inexpensive additive printed structures on highly flexible substrates and corrects the material and physical deformities by an adaptive integrated \ac{DBS} processor to generate fast autonomous insights. Specifically, we propose (1) a low-cost stable \ac{CuMOD} with $<0.1\%/^\circ$C resistivity variation, (2) energy-efficient beamformer ICs integrated on flexible substrates and capable of handling reduced supply margins to tackle the high resistivity of such printed inks, (3) real-time deformation correction with \ac{DBS} control of integrated phase-shifter and delay lines, and (4) rapid insight generation with area-efficient on-chip real-time \ac{DBS} processor.  

To the best of our knowledge, this is the first silicon-based additive printed flexible array addressing both physical and material deformities with a closed-loop \ac{DBS} processor. The proposed solution does not have high computational complexity and stabilizes the coherent signal output, supporting multi-point curvatures on non-planar shapes. The proven efficacy of this proof-of-concept prototype opens future pathways to autonomous algorithms for large-scale flexible arrays deployed in remote locations at the network edge.

\textcolor{black}{\section{Results}}
\label{results}
\textcolor{black}{This section outlines the performance of the \ac{DBS}-based, additively printed flexible antenna array (DBS-FLEX). The first subsection presents the performance of the \ac{CuMOD} ink, while the subsequent subsection discusses the overall array performance, including its \ac{DBS} capabilities.}\\

\subsection{Copper Molecular Decomposition Ink Properties}\label{Ink_results}
We characterize the ink stability and signal transmission under varying temperatures, corrosive environments, and mechanical bending reflected during practical use. \textcolor{black}{This enables us to evaluate further the long-term reliability of the proposed ink.  \textcolor{black}{ Herein, a molecular copper formate-based ink is utilized to create a highly conductive CuMOD ($=$35 MS/m)~\cite{inK_results} on a Pyralux substrate with consistent electrical and RF performance under varied strains and temperatures. Further, our recent experiments have shown that a secondary high-temperature annealing process can increase the conductivity up to 47 MS/m~\cite{Abdullah_ink}}. This ensures superior signal transmission and reduced energy loss, making it particularly suitable for high-frequency electronic devices. \ac{CuMOD} ink enables the fabrication of ultra-thin copper films with an average thickness of 250 nm, resulting in dense, continuous structures with high crystallinity and reduced electron scattering, thereby enhancing electrical performance. Unlike traditional methods that rely on precious metals such as silver and gold, our copper-based ink is cost-effective and sustainable, utilizing abundant materials and minimizing toxic chemical waste to address environmental concerns. The \ac{CuMOD}-based conductors also demonstrate exceptional mechanical stability under bending conditions and retain stable performance during prolonged exposure to elevated temperatures. This durability is crucial for flexible and wearable RF devices.} 

\begin{figure}[t!]
\centering
\vspace{-3mm}
\includegraphics[width=110mm]{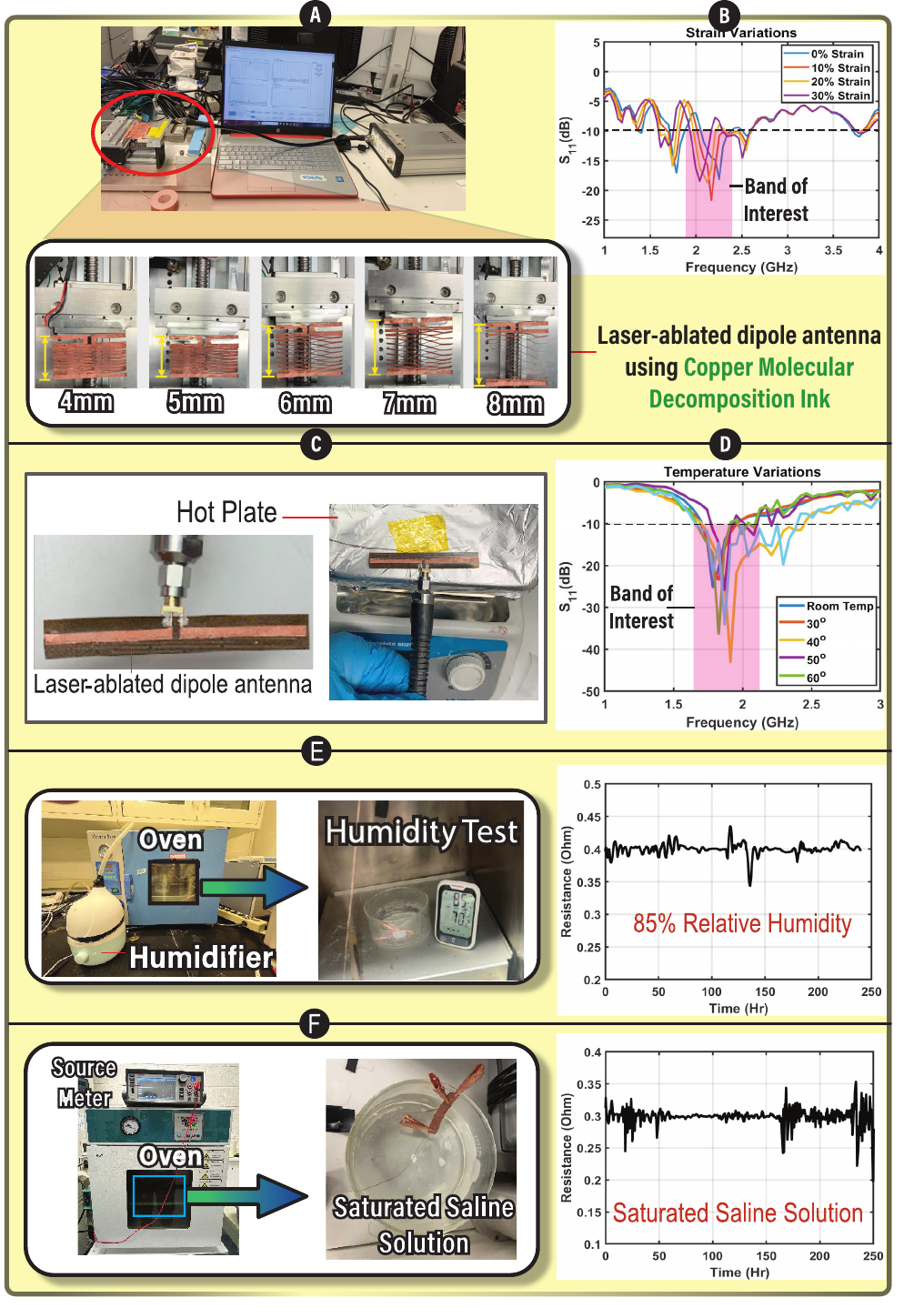}
\caption{\small \textcolor{black}{\textbf{a.} Laser-ablated dipole antenna printed using the proposed CuMOD ink testing under varied strains. \textbf{b.}  Measured $S_\text{11}$ with strain variations validates ink stability. \textbf{c.} Laser-ablated dipole antenna tested at different temperatures on a hot plate. \textbf{d.} Measured $S_\text{11}$ at different temperatures demonstrates ink stability. \textcolor{black}{\textbf{e.} Experimental setup to measure resistance variation over time of the CuMOD ink in an environment with 85\% relative humidity. \textbf{f.} Resistance of CuMOD ink in saturated saline solution measured over time (sample placed in a plastic-wrap sealed petri-dish containing saturated saline solution in a temperature-controlled oven at $25^\circ$C).  }} }
\vspace{-3mm}
\label{fig:fig2}
\end{figure}

\textcolor{black}{Figure~\ref{fig:fig2}a shows the test setup for validating the stability and RF properties of the ink under different strains. A simple stretchable dipole antenna is printed using the proposed ink, and the reflection coefficient, S\textsubscript{11}, is measured under strains varying from 4 mm to 8 mm. As shown in Fig.~\ref{fig:fig2}b, the S\textsubscript{11} is below 13 dB at the frequency of interest in every case, implying $>95\%$ signal transmission with very low reflection under strain. Similarly, the hot plate shown in Fig.~\ref{fig:fig2}c is used to measure the RF characteristics of the ink with temperature variations. The S\textsubscript{11} is captured for temperatures from 30$^\circ$ C to 60$^\circ$ C as shown in Fig.~\ref{fig:fig2}d. The values are well below -20 dB, which shows that more than 99\% of the signal is transmitted even with $50^\circ \text{C}$ change in temperature.} 

\begin{figure}[t]
\centering
\vspace{-3mm}
\includegraphics[width=105mm]{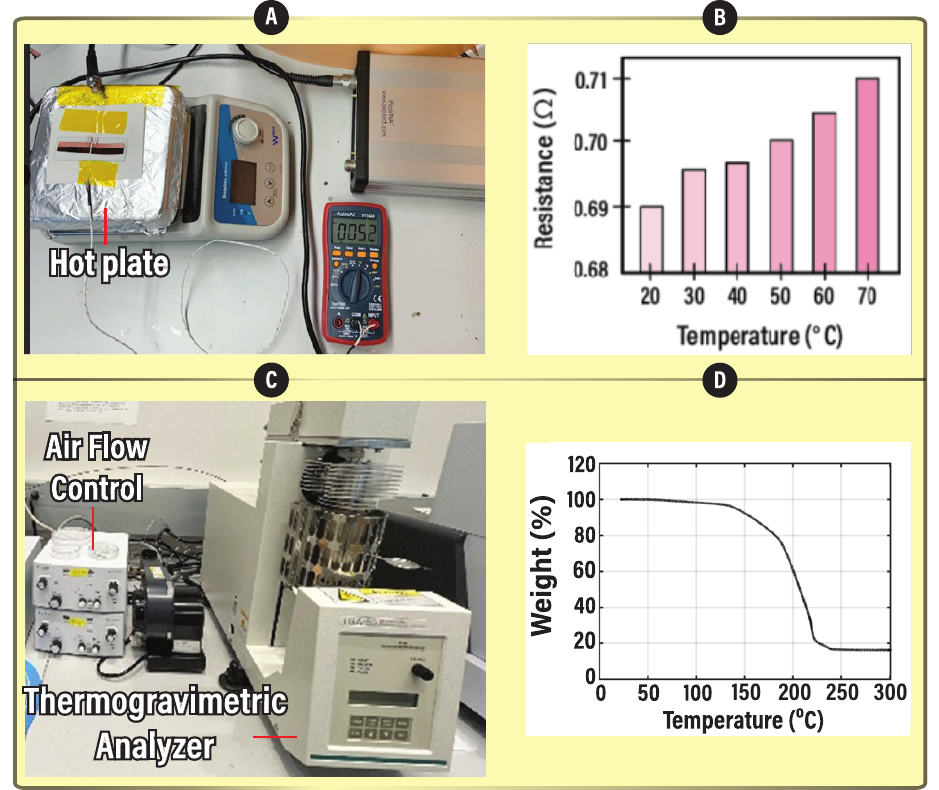}
\caption{\textbf{a.} Test setup for resistance measurements at different temperatures. Each printed trace using the \ac{CuMOD} ink is placed on a hot plate, and the resistance variation is captured against temperature variations. \textbf{b.} Measured resistance at different temperatures. \textcolor{black}{\textbf{c.} The experiment setup for \ac{TGA}, showing a microgram-sensitive balance, a controlled heating furnace, and a gas flow controller. \textbf{d.} The \ac{TGA} curve indicates that the decomposition temperature of the CuMOD ink is approximately $200–220^\circ C$.} }
\label{fig:TGA}
\end{figure}

\textcolor{black}{Additionally, we conducted tests to validate the ink’s stability in humid and saline environments, as shown in Fig.~\ref{fig:fig2}e and f. For reliability verification under humid conditions, the sample was placed in an oven, and water vapor was introduced using a humidifier to maintain a humidity level of approximately 85\%. The resistance of the sample was then measured using the four-point probe method. 
Figure~\ref{fig:fig2}e shows that the sample resistance remained stable over time under humid conditions. Figure~\ref{fig:fig2}f illustrates the ink's stability in a saturated saline solution. The sample was placed in a petri dish containing the solution, with half of the sample submerged. The petri dish was sealed with plastic wrap and tape to prevent water evaporation and was then placed in an oven maintained at $25^\circ$C. The resistance of the sample was measured using the two-point probe method with a Keithley 2450 SourceMeter. Figure~\ref{fig:TGA}a shows the test setup for validating the electrical properties of the ink at different temperatures. A simple DC line was printed, and a hotplate was used to observe resistance variations with respect to temperature. The ink remained stable, showing only a 0.02~$\Omega$ change in resistance across a 50$^\circ$C temperature change. This performance is comparable to bulk copper, with the measured change in resistivity less than 0.1\%$^\circ$C, and is better than that of state-of-the-art silver-based inks. \ac{TGA} on the samples, shown in Fig.~\ref{fig:TGA}c, revealed that the decomposition temperature of the ink is approximately 200–220$^\circ$C.}


\textcolor{black}{ \subsection{Active Array Processing with DBS}}
 \label{DBS_performance}

\begin{figure}[t!]
\centering
\vspace{-3mm}
\includegraphics[width=125mm]{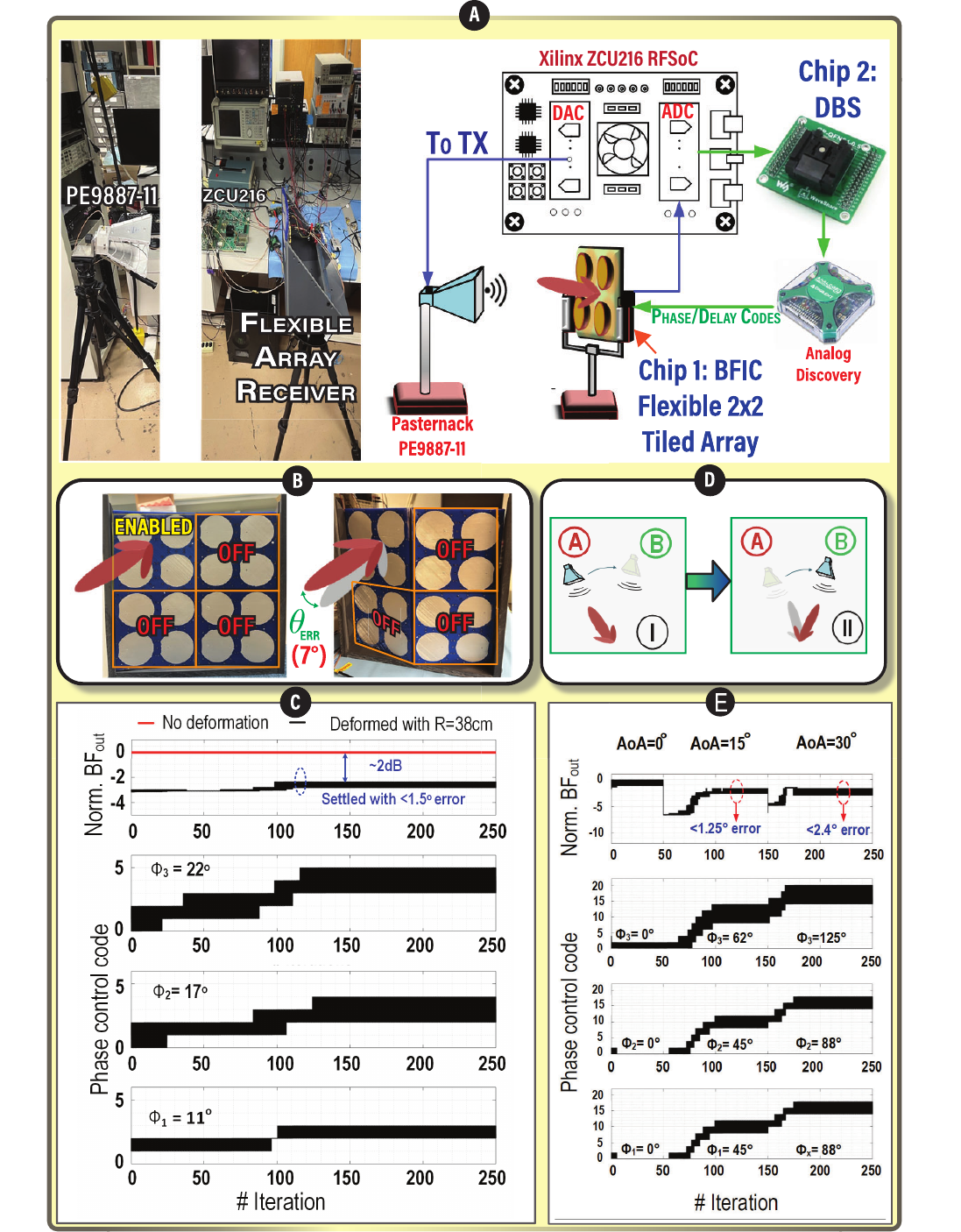}
\caption{\small Over-the-air measurement setup for deformation correction verification and radiation pattern.  \textbf{a.} Illustration of the test setup \textbf{b.} Worst-case deformation with 38 cm radius of curvature by tiling four flexible $2\times 2$ sub-arrays. \textbf{c.} The stabilization loop reduces deformation error to $<1.5^\circ$  from $7^\circ$; corresponding change in \ac{PS} codes $\phi_1$, $\phi_2$, and $\phi_3$   are captured. \textcolor{black}{\textbf{d.} Test setup illustration for dynamic phase adjustment verification by moving the transmitter. \textbf{e.} Dynamic phase adjustment validation when the transmitter moved from $0^\circ$ to $15^\circ$, and from $15^\circ$ to $30^\circ$.}  }
\label{fig:TB_Defr_pattern}
\end{figure}

\textcolor{black}{Figure~\ref{fig:TB_Defr_pattern} shows the DBS-FLEX array and the measurement setup used to validate the dynamic beam stabilization capabilities. Each 2$\times$2 antenna group, referred to as a tile, is connected to a \ac{BFIC}, which performs beamforming and \ac{DBS}. The beamformer is realized in the baseband using a discrete-time sample-and-hold architecture. The beamforming network consists of a time-interleaved switched-capacitor array that samples signals from each channel. Based on the \ac{AoA} and the delay experienced by each element, the sampling instances can be adjusted to extract coherent samples from each channel. The \ac{BFIC} output is fed to the \ac{DBS} through an off-chip \ac{ADC}, which controls the phase shift of each element to dynamically stabilize the beam under the conditions discussed in Section~\ref{sec:intro}.} The \ac{DBS} is synthesized digitally, facilitating integration into advanced fine-line CMOS technologies. Printing imperfections generate static offset errors from the predefined beam steering codebook. Dynamic deformations further alter the codewords for each antenna element from the corresponding beam steering angle, necessitating \ac{DBS}. To validate DBS-FLEX, we deform the antenna (noting that radial and circumferential deformations are identical due to the square structure of the tile) under different radii of curvature.

\textcolor{black}{As shown in Fig.~\ref{fig:TB_Defr_pattern}a and b, the 4$\times$4 tiled array is mounted on a custom antenna holder, enabling only one tile. All phase codewords were initially set to 1. The sides of the antenna holder were tightened to deform the array, as seen in Fig.~\ref{fig:TB_Defr_pattern}b. The minimum bending radius of curvature supported by our experimental setup is 38 cm, which results in a dynamic error of $7^\circ$. The $BF_\text{out}$ initially drops from its optimal value, and the phase codewords are automatically updated based on the gradient. Eventually, all phase codewords converge to their optimal values, and $BF_\text{out}$ reaches its optimum value, as shown in Fig.~\ref{fig:TB_Defr_pattern}c.}    \textcolor{black}{Each loop converges to minimize the beam pointing error $<1.5^\circ$. This demonstrates that during operation, the beam pointing errors induced by printing imperfections, initial placement, and dynamic deformation are compensated using DBS to maximize the \ac{SNR}.  This error can be further reduced by increasing the \ac{PS} resolution. The 2 dB gain reduction observed after beam-pointing error correction is attributed to the antenna holder and the change in Line-of-Sight (LOS). The phase control codes are properly initialized to prevent DBS from becoming stuck at side-lobes (local maxima).  }
\textcolor{black}{Because our experimental setup does not support a smaller radius of curvature ($<$38cm) or non-uniform multi-curvature bending, we further validated the dynamic phase adjustment capability by moving the transmitter from $0^\circ$ to $15^\circ$, and then from $15^\circ$ to $30^\circ$ to emulate dynamic changes. As shown in Fig.~\ref{fig:TB_Defr_pattern}d and Fig.~\ref{fig:TB_Defr_pattern}e, the phase codewords automatically adjust to their optimal values, validating the dynamic phase adjustment capabilities.} 

\begin{figure}[t!]
\centering
\vspace{-3mm}
\includegraphics[width=115mm]{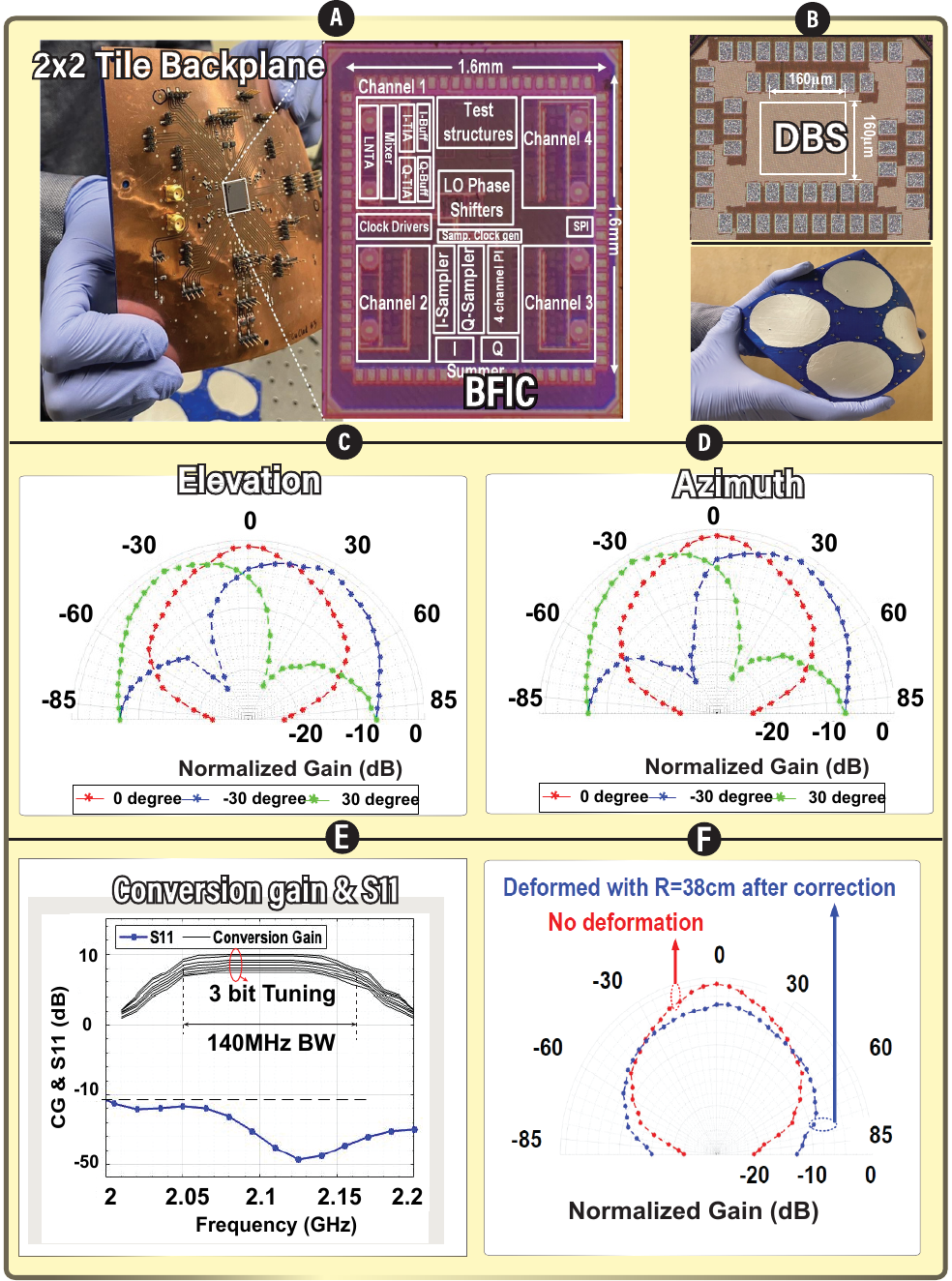}
\caption{\textbf{a.} The  $2 \times 2$ flexible array tile back side, BFIC, and its connections are printed using CuMOD ink on an AP layer, chip micrograph of BFIC.  \textbf{b.} Chip micrograph of DBS and front side of $2 \times 2$ flexible array with circular patch antenna printed on Ninjaflex.  \textbf{c.} Elevation and \textbf{d.} Azimuth radiation patterns of the $2 \times 2$ flexible array. \textbf{e.} S\textsubscript{11} and conversion gain of the single-receiver channel. \textbf{f.}  Radiation pattern without deformation and with deformation after correction at $0^\circ$ Azimuth.}
\label{fig:fig3}
\end{figure}

\textcolor{black}{Figure~\ref{fig:fig3}a shows the $2\times2$ tile backplane with additively printed CuMOD traces and the \ac{BFIC}. The chip micrograph of the \ac{BFIC} is shown in the inset, occupying only $1.6$ mm $\times~1.6$ mm of silicon area. Figure~\ref{fig:fig3}b (top) presents the chip micrograph of the \ac{DBS}, with an active silicon area of $160~\mu$m $\times~160~\mu$m. Figure~\ref{fig:fig3}b (bottom) shows the front side of the $2\times 2$ tile with circular antennas printed on a NinjaFlex substrate. The $2 \times 2$ conformal tile azimuth and elevation patterns under no deformation are shown in Fig.~\ref{fig:fig3}c and Fig.~\ref{fig:fig3}d. The measured BFIC single channel S\textsubscript{11} and conversion gain are shown in  Fig.~\ref{fig:fig3}e. The return loss is $<-10$ dB in the band of interest, and the 3-bit tunability in the BFIC enables gain calibration for each element. Figure~\ref{fig:fig3}f shows the radiation pattern without deformation and with deformation after applying \ac{DBS} (in the steady-state), highlighting the \ac{FoV} enhancement with minimal beam error under deformation (additional details on the BFIC performances are provided in Supplement~1.3). } The proposed DBS-FLEX array is lightweight and easily deployable with an areal mass of 0.464 $\text{g/cm}^2$ and a thickness of approximately 8 mm.


\textcolor{black}{\section{Discussions}}
\label{discussions}
\textcolor{black}{\textbf{CuMOD Ink:} Unlike prior art in gold, silver, carbon nanotubes, liquid metal, and conductive polymers, \ac{CuMOD} ink significantly closes the conductivity gap (see comparison in Supplementary~1.4 ) with bulk metals. As explained in Sec.~\ref{Ink_results}, experimental results indicate that \ac{CuMOD} ink exhibits an electrical conductivity up to $47 MS/m$. This is approximately 81\% of that of bulk copper, ensuring efficient signal transmission and reduced energy loss in high-frequency electronic devices. \ac{CuMOD} ink also excels in other key characteristics such as scalability, RF performance, additive manufacturability, and cost effectiveness~\cite{Abdullah_ink}.  Additionally, the \ac{CuMOD} ink demonstrates exceptional thermal stability and mechanical durability, making it highly suitable for applications in flexible electronic devices. \ac{CuMOD} ink is compatible with various additive manufacturing techniques, including inkjet printing, aerosol jet printing, and extrusion printing, significantly enhancing processing flexibility and scalability. Furthermore, its \ac{EMI} shielding effectiveness reaches up to 68 dB, making it ideal for wireless communication and other electronic applications, providing superior signal protection. Compared to noble metal-based conductive materials, \ac{CuMOD} ink offers a more cost-effective alternative and employs additive manufacturing processes, aligning with sustainable manufacturing trends and facilitating large-scale production and industrial applications.}


\textcolor{black}{\textbf{DBS-FLEX Power Efficiency:} As mentioned in Sec.~\ref{sec:intro}, beamforming is implemented with energy-efficient switched capacitor circuits. For algorithmic stabilization implementations, which are adopted by most state-of-the-art stabilization techniques, total power consumption is estimated based on the number of additions, multiplications, and memory accesses. Due to its low computational complexity, \ac{DBS} uses minimal addition and multiplication resources (see supplementary Table~1 for resource allocation of \ac{DBS}), and does not require an additional memory interface. The \ac{LUT} is implemented using registers since it requires very few perturbation levels. The multi-dimensional variable dynamic range search algorithm  in~\cite{Hajimiri_JSSC21} requires reprogramming of all phase settings in each iteration and also needs to run multiple batch processes in parallel. This demands high computational power and a memory interface. Furthermore, the additional recovery and generation units also consume power, leading to even higher power consumption. On the other hand, deep-learning-based correction presented in \cite{TAP23_DL} requires a memory interface to store the weights of each layer and needs to run multiple operations in parallel to find the optimal neural path, making it comparatively power-hungry. The lower computational complexity and power-efficient CMOS implementation make the proposed \ac{DBS} solution the most energy-efficient among state-of-the-art techniques. }

\begin{figure}[t!]
\centering
\vspace{-3mm}
\includegraphics[width=121mm]{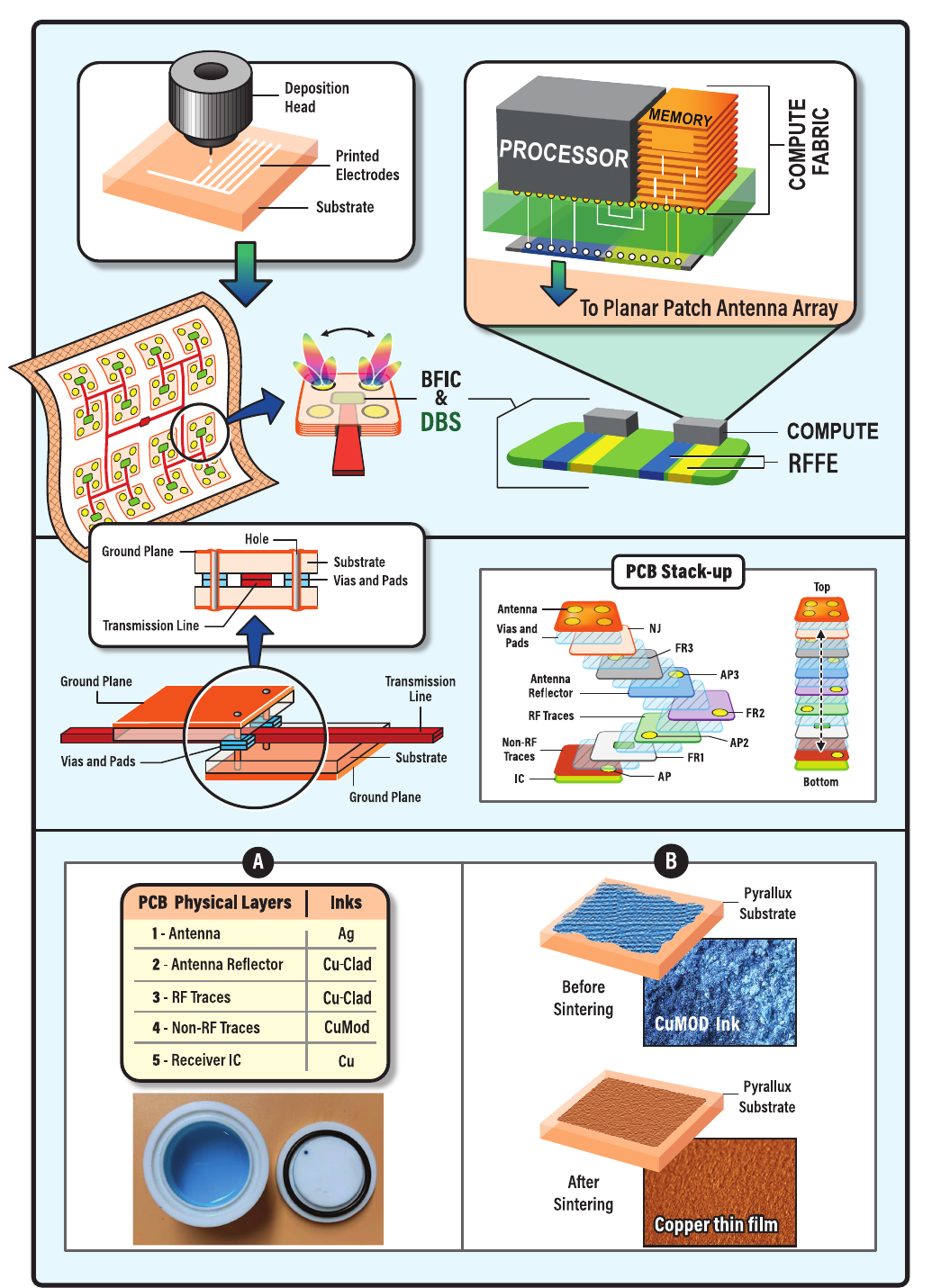}
\caption{Tile-based array scaling with integrated DBS processor attached to the additive printed flexible array. Future works on compute, communication, and sensing can be integrated at the network edge with fast insights and real-time BFIC adaptation. (Middle) Additively printed DBS-FLEX array stackup with NJ- NinjaFlex, FR- FastRase EZ, and AP- DuPont Pyralux. \textbf{a.} Full breakdown of inks in each array layer with ball-milled copper ink.  \textbf{b.} Scanning Electron Microscope (SEM) image of the ink before and after sintering. }
\label{fig:ink_strct}
\end{figure}

\textcolor{black}{\textbf{DBS-FLEX Scaling:} For large scale applications, tile-based realization of \ac{DBS} enables easy scaling to larger arrays~\cite{Tiled_scaling1}. Based on the antenna size at 2.1 GHz, we choose a $2 \times 2$ tile configuration. In terms of printing, as shown in Fig.~\ref{fig:TB_Defr_pattern}b, each tile is thermally bonded to its neighboring tiles. The tile-based structure alleviates the challenges associated with large-scale printing, which increases cost and imperfections, by allowing smaller tiles to be printed and bonded together to form a larger array. As shown in Fig.~\ref{fig:ink_strct}, multiple tiles, each with its own \ac{BFIC} and \ac{DBS}, are connected to scale the array. Since each \ac{DBS} only needs to optimize the phase codewords within its tile, it does not increase latency. Instead, multiple \ac{DBS} units can work in parallel to quickly obtain the optimal $BF_{out}$. The second-stage combiner, where all the tiles are coherently combined, requires a tile synchronization system to ensure that all tile-level optima contribute toward the global optimum in the scaled array. The discrete-time beamformer implementation of the \ac{BFIC} simplifies this task, as synchronization is achieved by aligning the clock signals fed to each tile. Furthermore, tile-based \ac{DBS} processing does not increase computational complexity as the array scales, since each tile is only responsible for optimizing the phase codes within its own tile, and convergence speed improves due to the inherent parallelism.}

\textcolor{black}{\section{Methods}}
\label{methods}
\subsection{Low-cost, reliable molecular copper decomposition ink }

Recent works on additive printing techniques such as aerosol/ink jetting, direct writing, and screen printing have shown enhanced scalability, reduced material waste generation, and cost-effective manufacturing (more details in Supplement~1.1)\cite{printing_review}. However, as highlighted in Section~\ref{sec:intro}, common challenges in additive manufacturing revolve around the printing method and electrical performance of the conductive inks. Factors such as additives, ink viscosity, uniformity, micro/nanomaterial structure, and size directly influence the potential printing method, substrate material, sintering conditions, and electrical performance of the traces.





Current ink developments drive a cheaper and higher-performance alternative to existing metallic, polymer, and carbon-based inks that are increasingly costly or fail to meet the electrical performance standards achieved through conventional manufacturing methods \cite{Yu23}. When considering conductive inks, metallic-based inks consistently demonstrate superior performance in RF electronics due to their inherently higher electrical conductivity, thermal conductivity, and suitability for various additive manufacturing methods \cite{silver_ink_variation}. Although silver (Ag)-based inks have been largely used due to their low resistivity (1.59 $\mu \Omega.cm$), copper-based inks are preferred due to their lower costs with comparable resistivity performance (1.72 $\mu \Omega.cm$) \cite{ink1}. However, excessive additives needed to improve the printability of copper nanostructured inks and with a greater potential for ambient condition oxidation suggest that molecular Cu-based inks are more advantageous for long-term commercial and industrial applications \cite{ inK_results,ink4}. 
The CuMOD ink material was made from a copper formate-based ink slurry developed through ball milling. \textcolor{black}{Copper formate was purchased from Thermo Fisher, while diethylene glycol butyl ether (DEGBE) and dimethylformamide (DMF) were obtained from Sigma. The slurry composition} followed a mass ratio of 0.45:5:0.45 for copper formate, DEGBE, and DMF, respectively. The copper ink synthesis was carried out in a ceramic ball milling container at a rotational speed of 300 rpm for 1 hour. The respective slurry was collected in a centrifuge tube after the mechano-chemical synthesis and centrifuged at 6000 rpm for 5 minutes and decanted to remove excess solvents. It was consequently redispersed in 2 mL of DEGBE and centrifuged and decanted once more to remove any excess DMF in solution. Figure~\ref{fig:ink_strct}b shows the ink before and after sintering. Before sintering, the ink is made of an approximate Cu formate flake size distribution from 1-25 $\mu m$. After sintering, the copper formate flakes reduce and form a densely percolated thin film of copper nanoparticles.



%

 \begin{figure}[t!]
\centering
\vspace{-3mm}
\includegraphics[width=125mm]{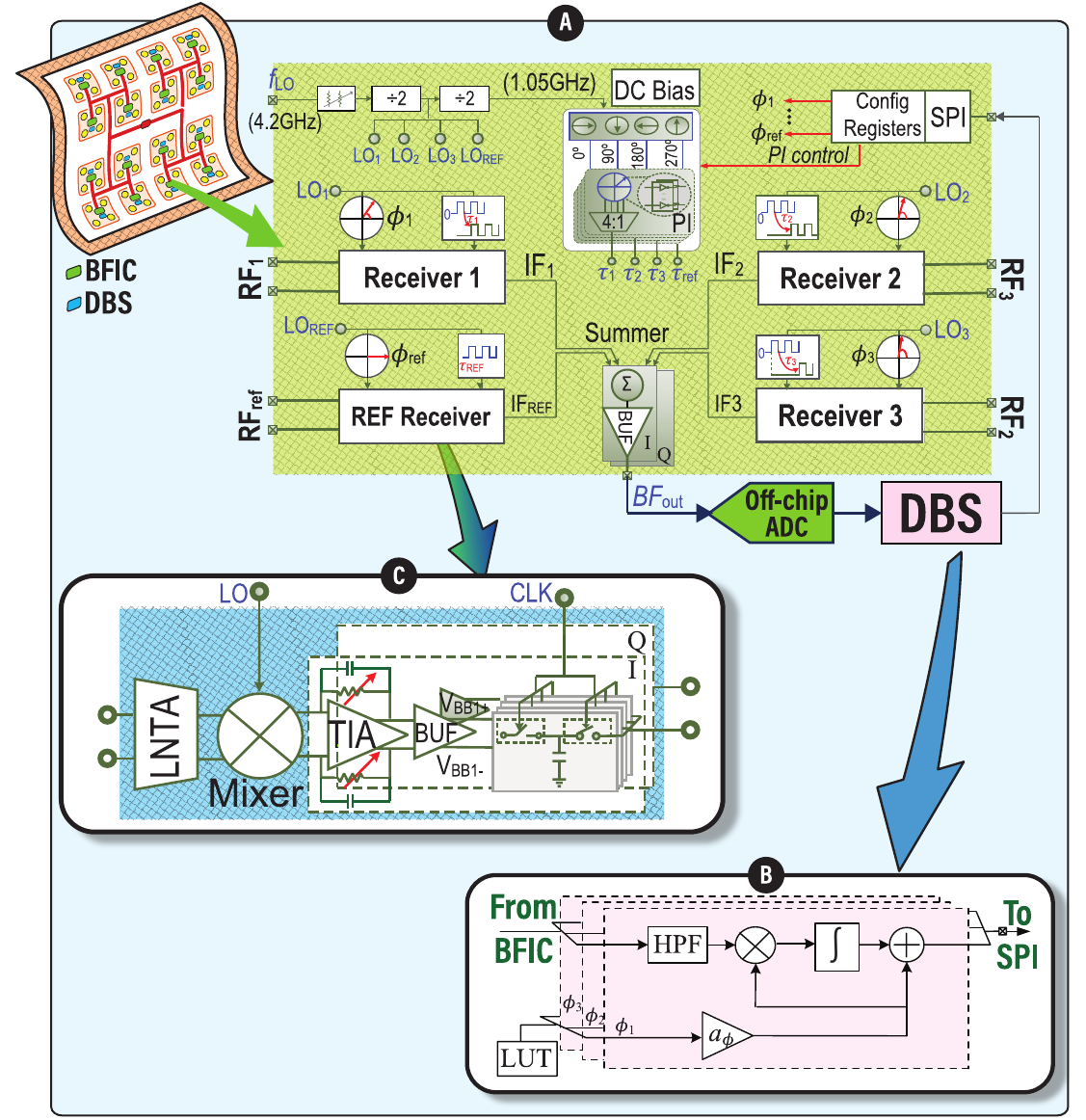}
\caption{\small \textbf{a.} 4-element beamforming receiver architecture with four downconverters and discrete-time beamformer with self-corrected stabilization loop in feedback. \textbf{b.} \ac{DBS} for correcting beam pointing errors, where HPF denotes a high-pass filter and LUT stands for look-up table. \textbf{c.} Single-channel receiver architecture with \ac{LNTA}, downconversion mixer, phase shifter, \ac{TIA}, and time-delayed sampler. Buffer (BUF), local oscillator (LO), intermediate frequency (IF), and serial-peripheral interface (SPI).} 
\vspace{-3mm}
\label{fig:RX_architecture}
\end{figure}

\textcolor{black}{\subsection{Silicon-based dynamic beam stabilization method  for a low-power active array under physical and material deformation}}
\label{subsec:beamStab}

  \begin{figure}[t]
        \centering
        \includegraphics[width=1.0\columnwidth]{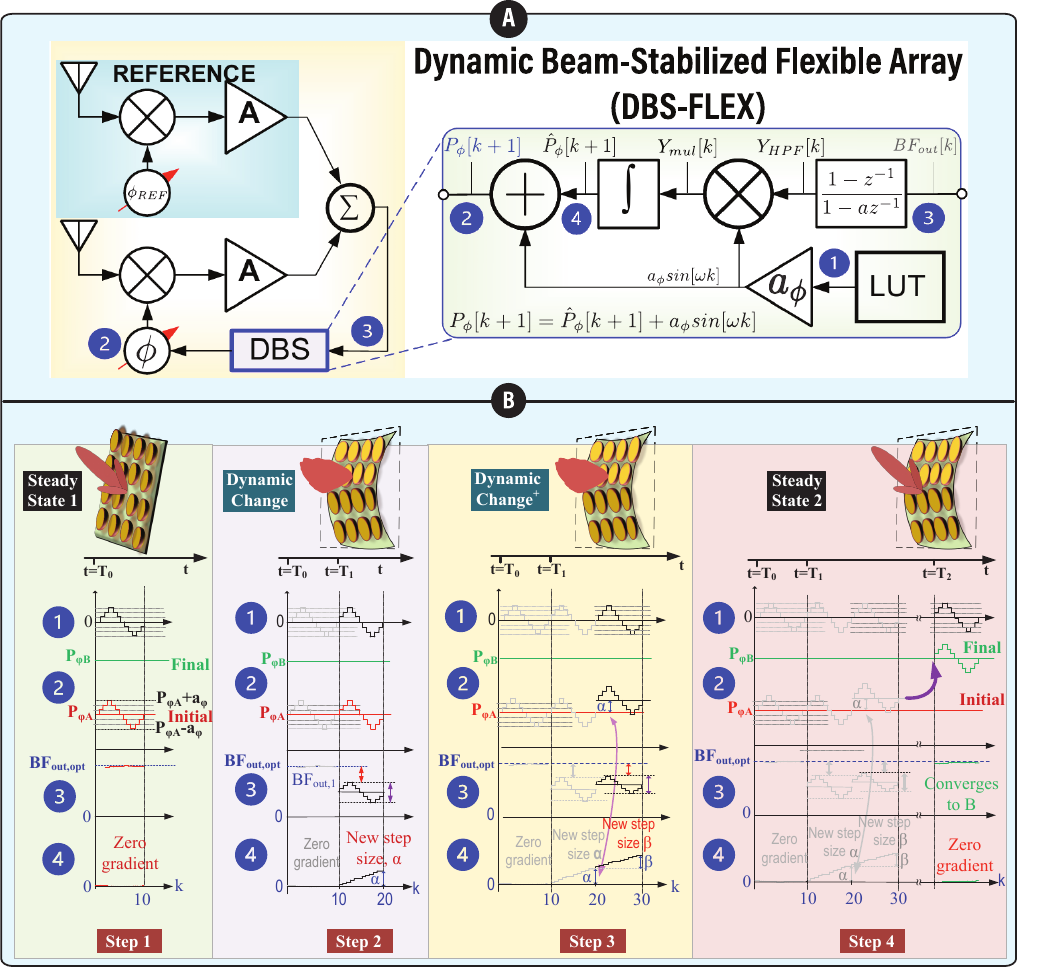}
        \vspace{-4mm}
        \caption{\small  \textcolor{black}{\textbf{a.} A block level representation of the \ac{DBS}. \textbf{b.} Illustration of automatic phase adjustment under dynamic changes. $P_{\phi A}$ is intial code word and $P_{\phi B}$ is  final phase codeword.} }
        \vspace{-4mm}
        \label{fig:DBS_dynamic_change}
    \end{figure}


As explained in  Section~\ref{sec:intro}, the additive printed conformal array is subjected to beam pointing errors during operation. These errors are caused by dynamic deformation and printing imperfections, which can lead to a loss in communication. Even a small misalignment can drastically degrade the \ac{SNR}. To address this issue, we introduce a beam stabilization loop into the silicon-based array processing for the flexible array. Given the power budget of the additive printed array, this proposed beam stabilization technique is low-power and low-area, inspired by model-free techniques such as perturb and observe, and extremum-seeking~\cite{ES2,ES3} principles.

As shown in Figure~\ref{fig:RX_architecture}a, the output of the active array IC, BF\textsubscript{out}, is fed into the beam stabilization loop because a beamformer has only one maximum point, which corresponds to the main lobe. The side lobes are considered local maxima. For each angle, corresponding to each deformation condition (single- or multiple-curvature), there is a unique phase shift combination (and corresponding phase code word) for each element. For an 8-bit phase shifter, there are thus $2^8$ possible phase shifts. At a given \ac{AoA} and a given deformation (single- or multiple-curvature), for a $2 \times 2$  array, there are $2^{8\times3}$ phase-shifting options, where one element is taken as reference (labeled as REF receiver in Fig.~\ref{fig:RX_architecture}a). Only one combination among these gives the maximum \ac{SNR} when the beam is perfectly aligned with the transmitter. Incremental time-based search counters to validate each option are impractical for the aforementioned applications targeting high mobility and lower latencies. A data-driven or model-driven approach using machine learning can provide the optimum combination with low latency, but is potentially susceptible to being stuck in an unknown condition, considering the plethora of uncertainties in real-world applications.  The proposed integrated loop ensures fast convergence to the optimum phase shifter combination, overcoming the above challenges.

\textcolor{black}{Figure~\ref{fig:DBS_dynamic_change}a shows the DBS-FLEX control loop. The beamforming output $BF_\textsubscript{out}$ from the \ac{BFIC} \Circled{3} is first fed into a \ac{HPF} to remove its mean. The HPF output is multiplied by a sinusoidal perturbation \Circled{1} from a \ac{LUT} and then integrated to estimate the gradient and the next step size \Circled{4}. The next step size is added to the sinusoidal perturbation and fed back \Circled{2} to the \ac{BFIC}.}
\textcolor{black}{Figure~\ref{fig:DBS_dynamic_change}b illustrates the step-by-step operation of the \ac{DBS} module for determining the optimal phase control code, $P_{\phi\text{,opt}}$ under a dynamic change. For simplicity, we consider only 2-elements, considering the first as \ac{REF} element as shown in Fig.~\ref{fig:DBS_dynamic_change}bw. For better conceptual illustration, we assume a very fine-resolution phase shifter.
\begin{enumerate}[label=    Step \arabic*:]
    \item At time t=$T_0$, the array is in a steady state with the phase shifter settled at $P_{\phi\text{,A}}$, providing an optimum beamforming output $BF_\text{out,opt}$. The \ac{LUT} signal \Circled{1} The \ac{LUT} signal \Circled{1} perturbs the phase shift around $P_{\phi\text{,A}}$. Since the system is in a steady state and the phase shift is at its maximum, the gradient is zero. Consequently, the next step size is also zero, and the system remains at $P_{\phi\text{,A}}$. 
    \item At time t=$T_1$, a dynamic change deforms the array, shifting the optimal phase code word to $P_{\phi\text{,B}}$. As a result, $BF_\text{out}$ drops from its optimal value, and the multiplier-integrator combination estimates the new step size $\alpha$ based on the gradient.
    \item The step size $\alpha$ estimated in the previous step is added to the phase code word, and the process is repeated. Based on the gradient at $P_{\phi\text{,A}}+\alpha$, a new step size $\beta$ is estimated. 
    \item After recursive iterations, the phase code word settles at $P_{\phi\text{,B}}$, resulting in the optimal $BF_\text{out,opt}$. This means the error induced by the dynamic change has been corrected by adjusting the phase code word. Since the phase is now at its optimum, both the gradient and step size will be zero.
\end{enumerate}}
\textcolor{black}{In this case $P_{\phi\text{,B}}>$ $P_{\phi\text{,A}}$. Supplementary Section 1.8 presents the mathematical framework illustrating the operation of each block in the \ac{DBS}, and further provides a detailed explanation of how \ac{DBS} performs gradient estimation using only array-level information, without element-level data.}

\textcolor{black}{\subsection{Integration of the additive printed array with flexible antennas and beam-stabilized active signal processing}} \label{subsec:integration}


Each sub-array \ac{BFIC} implements discrete-time true-time-delay-based signal combining presented by the authors in \cite{PI_TMTT,PI_TTD}, with this work showing the antenna-to-bits solution including the RF front-end and the discrete-time combiner with the entire clock generation (Fig.~\ref{fig:RX_architecture}a). The developed ink and the printing method ensure that the four channels of the \ac{BFIC} are connected to each antenna through impedance-matched via holes and traces. Each receiver front end provides a low-noise impedance matching to the antenna/PCB trace impedance. Each receiver is designed to support higher modulation schemes, including \textit{M}-QAM (Quadrature Amplitude Modulation) and OFDM (Orthogonal Frequency-Division Multiple Access). To process the higher modulated signals, the receivers down-convert the incoming RF signals to the baseband while extracting the in-phase (\textit{I}) and the quadrature-phase (\textit{Q}) components.  Beamforming of the four channels is performed in the baseband after extraction of the quadrature phases using a discrete-time beamformer \cite{PI_TTD}. Time-delayed sampling of the extracted quadrature baseband signals along with the \ac{LO} \ac{PS} captures coherent samples from each element and enables wideband beamforming as shown in  Fig.~\ref{fig:RX_architecture}a. Independent gain, phase, and delay tuning is provided for each of the quadrature channels to reduce the respective mismatches in each channel, leveraging DBS as shown in Section~\ref{subsec:beamStab}. An impedance-controlled high-frequency external clock ($2 \times $RF input frequency) is provided for the signal down-conversion and time-delayed sampling, which is susceptible to impedance variations due to deformation and printing imperfections. A tunable impedance matching network is used at the clock input to alleviate these impedance imperfections.

The PCB stack-up for the prototype flexible tile is also shown in Fig.~\ref{fig:ink_strct} with the 4-channel BFIC attached to a $2 \times 2$ additive printed array with the antenna also printed in the backplane of the PCB. The substrate comprises multiple flexible sheets of DuPont Pyralux\textsuperscript{\textregistered} AP (polyimide) and Ninjaflex. The BFIC, and its RF and non-RF traces, are printed over AP layers of thickness 0.127 mm combined using 0.0508 mm Fast Rise EZ (epoxy). The antenna ground plane is printed over 0.254 mm AP combined with other layers using 0.0508 mm Fast Rise EZ. The antennas have been printed on a Ninjaflex substrate with 7.62 mm thickness chosen to meet the ground plane height requirements at 2.1 GHz. Circular patch antennas are used, as bending has minimal impact on the return loss and gain. The three DuPont Pyralux\textsuperscript{\textregistered} AP sheets are combined using Fast Rise EZ as mentioned, which are then thermally bonded with the Ninjaflex substrate.\\
\indent \textcolor{black}{A complete breakdown of the inks used in each layer is shown in Fig.~\ref{fig:ink_strct}. The implemented \ac{BFIC} is located on layer 5, with connections printed using CuMOD ink. The non-RF traces on layer 4 are also printed with CuMOD ink. Commercial off-the-shelf (COTS) Cu-clad is used for the RF traces on layer 3 and the antenna reflectors on layer 2. Finally, Ag (silver) ink is used to print the antennas on layer 1. Silver ink was chosen for its ability to sinter at low temperatures, preventing warpage of the 3D-printed \ac{TPU} substrate. It also enabled direct-write deposition of the antennas onto the TPU substrate. Ag vias and Cu ink vias reinforced with Ag are used to connect between layers. The signal from each additively printed antenna on layer 1 is tapped through impedance-matched Cu–Ag via holes for maximum power transfer to layer 5. The measured printed trace and via performances are provided in Supplementary~1.6.}\\

\section{CONCLUSIONS}
The proposed DBS-FLEX arrays address critical shortcomings related to motion-induced deformation in arrays targeted for conformal surfaces. The fundamental limitations related to additive printing imperfections and dynamic deformation mitigation are overcome with multi-variable independent optimization demonstrated on a silicon-based active array processor that is scaling-friendly and suited to mitigate any printing, strain, and temperature-related deformations. This approach is particularly promising for applications like textile antennas and airborne platforms, which demand both flexibility and durability with low SWaP-C.

 \indent \textcolor{black}{Future works will investigate direct-printing methods with a lower-temperature sintering process with high conductivity that would enable direct printing on a wider variety of materials such as PET, \ac{TPU}, paper, and textiles, leveraging research into compounds such as amino-methanol.  Similarly, additional computational techniques need investigation for making \ac{DBS} versatile for arbitrary waveforms and spatial directions.}

\section{DATA AVAILABILITY}
The datasets generated and additional results are available from the corresponding author upon reasonable request. \\
\section{ACKNOWLEDGEMENTS}
\thanks{This material is based on research sponsored, in part, by Air Force Research Laboratory under agreement number FA8650-20-2-5506, as conducted through the flexible hybrid electronics manufacturing innovation institute, NextFlex, Murdock Foundation, Washington Research Foundation, and WSU Office of Commercialization. The U.S. Government is authorized to reproduce and distribute reprints for Governmental purposes, notwithstanding any copyright notation thereon.

The views and conclusions contained herein are those of the authors and should not be interpreted as necessarily representing the official policies or endorsements, either expressed or implied, of Air Force Research Laboratory or the U.S. Government.

The authors acknowledge Dr. Robert Dean at Auburn University for his help with the antenna hold and Sonja Gerard at OEIGraphics for help with Figures 1 and 2.}%

\section{AUTHOR CONTRIBUTIONS}
This project was conceived by S.P. and S.G.. The antenna was designed and additive printed by T.D. and K.K under the supervision of J.N., and J.W. The CuMOD ink was developed and verified by A.I. and S.R. \textcolor{black}{Z.W. performed additional experimental validations of the CuMOD ink with input from A.I. and S.R. } The receiver front-end was designed by S.P., A.M., and A.R. with inputs from S.S. and S.G. The phase-interpolator and time-interleaved clock generator were designed by S.P. with inputs from S.S. and S.G. The DBS loop, including the receiver front-end integration with the discrete-time beamformer, design, and validation, was done by S.B. and S.P. with inputs from S.G. Measurements of the DBS-FLEX array were performed by S.P. with inputs from S.G. S.P. and S.G. prepared the manuscript with comments from all other authors.

\section{COMPETING INTERESTS}
The authors declare no competing non-financial interests. 
\newpage

\newpage
\clearpage

\bibliography{sn-bibliography}

\end{document}